\documentclass{article}

\usepackage{arxiv}

\usepackage[utf8]{inputenc} 
\usepackage[T1]{fontenc}    
\usepackage{hyperref}       
\usepackage{url}            
\usepackage{booktabs}       
\usepackage{amsfonts,amsmath,amssymb}       
\usepackage{nicefrac}       
\usepackage{microtype}      
\usepackage{graphicx}
\usepackage{caption}

\usepackage{color, colortbl}

\usepackage{mathptmx,amsthm}
\usepackage{xcolor}
\usepackage{multirow}
\usepackage{subcaption}

\usepackage{natbib}

\newcommand{\ie}{\textit{i.e.,}}
\newcommand{\eg}{\textit{e.g.,}}
\newcommand{\commenttxt}[1]{}

\newcommand{\mybar}{\kern1pt\rule[-\dp\strutbox]{.8pt}{\baselineskip}\kern1pt}

\newcommand{\noo}{NO\textsubscript{2}}
\newcommand{\soo}{SO\textsubscript{2}}
\newcommand{\cmo}{CO}
\newcommand{\pmv}[1]{PM#1}
\newcommand{\ooo}{O\textsubscript{3}}
\newcommand{\caqv}{CityAQVis}
\newcommand{\gee}{GEE}
\newcommand{\ml}{ML}

\renewcommand{\mytagline}{} 

\theoremstyle{definition}

\title{\caqv: Integrated ML-Visualization Sandbox Tool for Pollutant Estimation in Urban Regions Using Multi-Source Data (Software Article)} 

\author{
  Brij Bidhin Desai,
  Yukta Arvind Rajapur,
  Aswathi Mundayatt,
  Jaya~Sreevalsan-Nair\thanks{\texttt{jnair@iiitb.ac.in}} 
  \\
  Graphics-Visualization-Computing Lab,\\
  International Institute of Information Technology Bangalore, Karnataka 560100, India. \\
  \texttt{http://www.iiitb.ac.in/gvcl} 
}

\begin{document}
\maketitle

\begin{abstract}
  Urban air pollution poses significant risks to public health, environmental sustainability, and policy planning. Effective air quality management requires predictive tools that can integrate diverse datasets and communicate complex spatial and temporal pollution patterns. There is a gap in interactive tools with seamless integration of forecasting and visualization of spatial distributions of air pollutant concentrations. We present CityAQVis, an interactive machine learning (\ml) sandbox tool designed to predict and visualize pollutant concentrations at the ground level using multi-source data, which includes satellite observations, meteorological parameters, population density, elevation, and nighttime lights. While traditional air quality visualization tools often lack forecasting capabilities, CityAQVis enables users to build and compare predictive models, visualizing the model outputs and offering insights into pollution dynamics at the ground level. The pilot implementation of the tool is tested through case studies predicting nitrogen dioxide (\noo) concentrations in metropolitan regions, highlighting its adaptability to various pollutants. Through an intuitive graphical user interface (GUI), the user can perform comparative visualizations of the spatial distribution of surface-level pollutant concentration in two different urban scenarios. Our results highlight the potential of \ml~driven visual analytics to improve situational awareness and support data-driven decision-making in air quality management.
\end{abstract}


\section{Introduction} \label{sec:introduction}
Air pollution poses a significant threat to public health and the environment, with pollutants such as \noo, particulate matter (\pmv{2.5} and \pmv{10}), and ozone (\ooo) contributing to a wide range of adverse health outcomes and environmental degradation. Monitoring and analyzing pollutant concentrations is critical for developing effective mitigation strategies and understanding urban air quality dynamics. Urban regions are of interest for air quality studies owing to their relationship with urban form, human mobility, intense socio-economic activity, and urban sprawl, and their impact on public health~\citep{hankey2017urban}.

Traditional methods of air quality monitoring, which rely heavily on ground-based sensors, are often limited by sparse spatial coverage and inconsistent temporal availability. Satellite-based approaches, such as those employed by \cite{wu2020space}, provide a scalable method for quantifying urban emissions, complementing ground-based monitoring stations. The advent of satellite remote sensing technologies, such as the TROPOspheric Monitoring Instrument (TROPOMI) aboard the Copernicus Sentinel-5 Precursor satellite, has enabled global monitoring of atmospheric pollutants with unprecedented spatial and temporal resolution. Integrating these datasets with meteorological, demographic, and land-use information provides a more holistic approach to understanding air pollution variability.

With the improved access to multi-source data, artificial intelligence (AI) and machine learning (\ml) models have emerged as powerful tools to predict air quality. These models use multi-source or multimodal datasets to uncover complex spatiotemporal patterns. However, the performance of these models often varies across pollutants and urban settings due to differences in emission sources, topography, and meteorological conditions. Hence, it is important to adaptively choose optimal models by comparing the outputs of multiple models across different cities, or even within a city across different time instances.

To address this, we introduce \caqv, a configurable \ml~sandbox tool that enables users to build, compare, and visualize predictive models for ground-level pollutant concentrations in urban regions. The tool uses multi-source datasets, including satellite-derived pollutant observations, meteorological parameters, population density, elevation, and nighttime lights as inputs, thus offering flexibility across different pollutants and urban regions.

\caqv~is demonstrated through case studies predicting \noo~concentrations in Delhi and Bangalore, two Indian cities with contrasting air quality profiles. While \noo~serves as the focal pollutant for these demonstrations, the underlying framework is pollutant-agnostic and can be readily extended to other air quality indicators. By allowing users to explore variations across pollutants, cities, and \ml~models, \caqv~supports data-driven decision making for urban air quality management.

The main contributions of this work are as follows:
\begin{itemize}
    \item \textbf{The design and implementation of \caqv, an integrated and interactive ML sandbox tool.} This tool addresses a gap in current resources by combining predictive modeling, data processing, and visualization into a single, user-friendly environment. It allows users to build, train, and compare different models in real time.

    \item \textbf{A configurable framework for urban air quality analysis that uses multi-source data.} The tool's architecture is designed to be pollutant-agnostic and easily adaptable to different cities and time periods. It effectively integrates diverse datasets, including satellite observations, meteorological parameters, population density, and nighttime lights, to provide a holistic approach to pollutant estimation.

    \item \textbf{Use of comparative visualization as a data-driven decision-making for urban air quality analysis.} \caqv~features an interactive GUI designed to make complex ML outputs accessible and interpretable. It empowers urban planners and researchers by enabling direct visual comparison of pollution patterns across different scenarios, such as between two cities or across different years for a single city.

\end{itemize}

\section{Related Work} \label{sec:related_work}
Though most of the studies provide visualization of outcomes, \caqv~is uniquely designed for users to explore datasets and analyze them with the help of comparative visualizations.

\subsection{Analysis Using Multi-Source Data}
Recent studies in air quality prediction and modeling have increasingly leveraged \ml~techniques and multi-source datasets to enhance the estimation and analysis of pollutant concentrations, particularly \noo~concentration. The TROPOspheric Monitoring Instrument (TROPOMI) onboard the Copernicus Sentinel-5 Precursor (S5P) satellite, launched by the European Space Agency in October 2017, entered its operational phase in April 2018, from which point data suitable for analysis became available. We observe that several data analytical studies align with the objectives of \caqv~by integrating TROPOMI satellite data, ground-based measurements, and ancillary information for improved urban air quality prediction. While several of these studies are predominantly focused on specific study areas, our proposed tool \caqv~is designed for comparative analysis across different study areas.

\subsubsection{Including TROPOMI Observations}
\cite{chan2021estimation} estimated surface \noo~concentrations over Germany using an artificial neural network model that was trained using TROPOMI observations of tropospheric \noo~and meteorological variables, and validated using ground observations. This work showed that the ML models outperformed the regional chemical transport models in analyzing the spatio-temporal variability of surface \noo, including changes due to the COVID-19 pandemic. Similarly, \cite{grzybowski2023estimations} compared different ML approaches, namely, linear regression, multiple linear regression, random forests, and radial kernel support vector machines, to estimate surface \noo~distributions over Poland, using meteorological, environmental, and socio-economic factors. This study concluded that the random forest model was the most effective, with the least error, and could be used for at least 120 days per year with cloud-free conditions. It also demonstrated that the TROPOMI \noo~column number density (TVCD) was the most significant factor or variable for these predictions. \cite{rowley2023predicting} expanded the study to fuse TROPOMI \noo~dataset with Sentinel-2 multispectral images, along with ground monitoring station properties, and train convolutional neural networks (CNNs), named AQNet, to predict one- or three-pollutant concentration. The outputs, which included concentrations of \noo, \ooo, and \pmv{10}, demonstrated that deep learning models and richer datasets enable air quality indexing in specific locations of interest. Similar studies were conducted in China during COVID-19 period~\citep{wang2021estimating,long2022estimating}, in California during 2018-2019~\citep{lee2023neighborhood}, in Central-East Europe during 2018-2022~\citep{wieczorek2023air}, and most of Europe during 2019-2021~\citep{shetty2024estimating}. The use of SHAP value analysis by~\cite{shetty2024estimating} again demonstrates the high significance of TROPOMI \noo~column density for surface \noo~estimation, but also provides an example of an explainable AI application.

There are methods outside the ML umbrella that use TROPOMI observations to build analytical models or provide insights into the multi-source datasets. A detailed analysis in Paris by~\cite{lorente2019quantification} arrived at a simple model of the processed TROPOMI \noo~line density as a function of the along-wind distance (meteorological data) exclusively over the city. The values simulated using the model were used to determine the strength and spatial distribution of city-wide emissions. A study conducted in Finland by~\cite{virta2023estimating} recently confirmed the high correlation of surface \noo~measurements with TROPOMI observations. Similar studies were conducted using ground measurements and TROPOMI data of \noo~in China during the COVID-19 period~\citep{wu2021spatiotemporal}, and South Korea in 2019~\citep{jeong2021assessment}.

Given its operational timeline, TROPOMI data have been critical for various applications involving air quality assessment during the COVID-19 pandemic in 2019-2020. For instance, the pollutant measurements from ground monitoring stations and TROPOMI showed high correlation to COVID-19 hotspots in fourteen cities in India in early 2020~\citep{naqvi2021spatio}.

\subsubsection{Without TROPOMI Data}
Different from surface \noo~estimation and without relying on TROPOMI satellite observations, \cite{ravindiran2023air} conducted a comparative analysis of multiple ML models trained on a diverse dataset comprising twelve air pollutants/contaminants and ten meteorological parameters to predict the Air Quality Index (AQI) in the city of Visakhapatnam. This study demonstrated that the CatBoost model outperformed other models, including Random Forest. A similar study by \cite{samad2023air} compared several ML regressor models for predictions of \pmv{2.5}, \pmv{10}, and \noo~levels using meteorological data, traffic information, and pollutant concentrations as measured in nearby monitoring stations in German cities. Their results support the viability of ML models in estimating pollutant concentrations without extensive ground sensor deployment, but reinforce the significance of information from nearby monitoring stations for accurate predictions.

Recently, \cite{rahman2024airnet} developed AirNet, a real-time web-based air quality forecasting platform using global pollutant data, such as \cmo, \ooo, \noo, and \pmv{2.5}, as features for a regression model. The model output gives a categorical rating of air quality for a given set of input values. Different from AirNet, \caqv~prioritizes accessibility and interactivity but goes further by offering a sandbox environment where users can build, compare, and visualize customized predictive models using a variety of ML algorithms. Also, \caqv~allows learning of a new model for localized data, as opposed to a global model used for any region in AirNet.

\subsection{Integrated ML-Visualization Systems}
\cite{vellido2011seeing} emphasized the crucial role of visualization in machine learning, not only as a tool for result presentation but also as a means to enhance interpretability and understanding, especially for non-expert users. \caqv~is hence designed to integrate visual analytics to bridge the gap between complex model outputs and actionable insights. Especially, urban and environmental studies have increasingly leveraged interdisciplinary tools for scalable analyses of heterogeneous datasets~\citep{grubert2016benefits}. Enhancing trust in \ml~systems is also a key consideration. \cite{chatzimparmpas2020state} argued that transparency through visualization and explanation is essential for fostering user trust. We use this as a design principle in \caqv~wherein the tool enables users to inspect model predictions interactively, thereby promoting transparency in urban air quality prediction using ML models. 

Lastly, the design concepts of the sandbox ML tools, as discussed by \cite{nodalo2019building}, serve as supporting guidelines for the development of \caqv. These guidelines highlight the main features of our platform, which are designed to democratize machine learning for monitoring urban air quality: flexibility, user control, and ease of model experimentation.

Open toolkits such as Google Earth Engine (\gee) flexibly allow users to search and load datasets, and perform visualization, exploration, ML model implementation, and analysis in one place~\citep{bahadur2023applications}. There are several studies which use \gee~for TROPOMI \noo-based analysis, as listed by~\cite{bahadur2023applications}. \caqv~also uses \gee~to access the data for driving factors in our ML models. While a similar land cover classification toolkit using GEE, O-LC Mapping by~\cite{xing2021lcmapping} provides a user interface, \caqv~provides functionality for combining several driving factors to estimate the \noo~concentration.

In information visualization, comparative visualization enables the cognitive task of comparing two different outcomes to evaluate the effectiveness of settings, \eg~ML model choice, input datasets, study area, etc. The design patterns for comparative visualizations include juxtaposition, which is widely used, and others, such as superimposition, explicit encoding of difference, and explicit encoding of time warp~\citep{gleicher2011visual}. The weakest or lazy coupling of visualizations in a juxtaposed view makes composing existing visualizations straightforward~\citep{javed2012exploring}. However, the juxtaposition design shifts the task of identifying the connection between the visualizations to the user of the visualization~\citep{gleicher2011visual}. Here, we address this challenge by allowing the user to select the settings for each of the two views, where the task of setting up the visualizations provides the user with an implicit understanding of the connection between the two views.

\section{Multi-Source Datasets of Interest} \label{sec:datasets}
Before describing the proposed tool, \caqv, it is important to get an impression of the datasets of interest for such a tool. We describe the datasets used in the case study to demonstrate the usage of the tool later in this article. Table~\ref{tab:data} gives a summary of the multi-source data used as input in the ML regression models in \caqv, and the output variables.

\begin{table}[tp]
  \centering
  \caption{A summary of input and output variables used in the ML models in \caqv}
  \label{tab:data}
  \renewcommand{\arraystretch}{1.3} 
  \begin{tabular}{>{\centering\arraybackslash}m{2.8cm} 
                  >{\centering\arraybackslash}m{2.5cm} 
                  >{\centering\arraybackslash}m{1.8cm} 
                  >{\centering\arraybackslash}m{1.2cm} 
                  >{\centering\arraybackslash}m{1.6cm} 
                  >{\centering\arraybackslash}m{1.0cm}}
    \hline
    \textbf{Variable Name} & \textbf{Data Source} & \textbf{Resolution} & \textbf{Unit} & \textbf{Frequency} & \textbf{Type} \\
    \hline
    Atmospheric NO$_2$ Column Density (TVCD) & TROPOMI (Sentinel-5P) & 3.5 × 5.5 km & mol/m$^2$ & Daily & Input \\
    Ground-level NO$_2$ Concentration & CCRAQM (CPCB) & Point-based & $\mu$g/m$^3$ & Monthly & Output \\
    Rainfall & JAXA GPM GSMaP v6 & $\sim$0.1° & mm/month & Monthly & Input \\
    Temperature & TerraClimate (IDAHO EPSCoR) & $\sim$4 km & Kelvin (K) & Monthly & Input \\
    Wind Speed & TerraClimate (IDAHO EPSCoR) & $\sim$4 km & m/s & Monthly & Input \\
    Population Density & GPWv4 (CIESIN/Columbia Univ.) & $\sim$1 km & persons/km$^2$ & Static & Input \\
    Elevation & CGIAR-CSI SRTM 90m & 90 m & meters (m) & Static & Input \\
    Nighttime Lights & NOAA VIIRS DNB VCMCFG v1 & 500 m & nW/cm$^2$/sr & Monthly & Input \\
    \hline
  \end{tabular}
\end{table}
\subsection{Study Areas Used in Case Study of \caqv}
Bangalore and Delhi, which are cities in India, were chosen as study sites for the demonstration of the tool usage in this article. These sites are urban regions of interest to environmental researchers due to their unique urban sprawls, leading to distinct air quality profiles and relevance to study ambient levels of \noo.

Bangalore represents a rapidly growing urban environment. Although experiencing rapid urbanization and an increase in vehicular emissions, Bangalore's air quality challenges are moderate compared to Delhi's. Its climatic conditions, influenced by higher altitudes and greenery, present unique variables for the prediction of \noo~\citep{mathew2024unveiling}.

On the other hand, Delhi, often ranked among the most polluted cities worldwide, experiences seasonal variations in the levels of \noo, \cmo, \ooo, and \soo, and these values tend to be high due to a combination of vehicular emissions, industrial activities, and unfavorable meteorological conditions~\citep{ansari2025seasonal}. It provides a robust test case for validating the ability of \caqv~to model extreme pollution scenarios and evaluate its responsiveness to varying urban dynamics.

\subsection{TROPOMI Tropospheric \noo~Column Density}
The TROPOspheric Monitoring Instrument (TROPOMI) is a passive, nadir-viewing, satellite-based push-broom imaging spectrometer aboard the Copernicus Sentinel-5 Precursor (S5P) satellite. It operates in a sun-synchronous orbit at an altitude of approximately 824 km, with a local equator crossing time of 13:30 during its ascending node. TROPOMI features eight spectral bands that span ultraviolet (UV), visible (VIS), near-infrared (NIR), and shortwave infrared (SWIR) wavelengths. It captures measurements at 450 positions across its orbital track, covering a swath of about 2600 km, thus enabling global daily coverage. Initially, the spatial resolution of the instrument was 3.6 km (across the track) by 7.2 km (along the track). On 6 August 2019, this resolution was improved to 3.6 km (across track) by 5.6 km (along track). In this study, we used the level-3 tropospheric vertical column density (TVCD) of $\text{NO}_2$, which is expressed in mol/m$^2$, and its 2019 yearly composite visualization is illustrated in Figures~\ref{fig:blr_drivingfactors}(a) and~\ref{fig:del_drivingfactors}(a).

The TROPOMI NO$_2$ dataset is publicly available via \gee~at: \url{https://developers.google.com/earth-engine/datasets/catalog/COPERNICUS_S5P_OFFL_L3_NO2}.

\subsection{In-situ \noo~Station Data}
\begin{figure}
    \centering
    \includegraphics[width=1.0\linewidth]{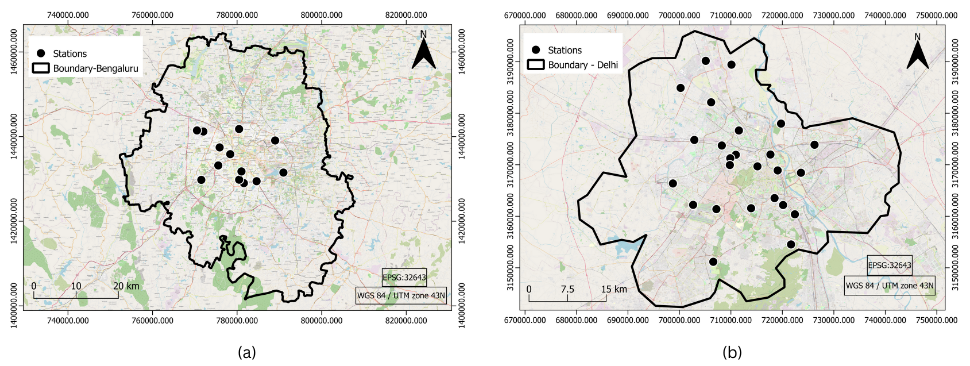}
    \caption{In-situ \noo~monitoring stations for selected cities in India, namely, (a) Bangalore, and (b) Delhi}
    \label{fig:stations}
\end{figure}

The \noo~surface concentration values, expressed in $\mu g/m^3$, for both Bangalore and Delhi, were obtained from the Central Control Room for Air Quality Management (CCRAQM) Portal. CCRAQM provides hourly and daily data for various pollutants. 13 stations were chosen from Bangalore, and 25 stations in and around Delhi. Their spatial distribution is depicted in Figure~\ref{fig:stations}.

The data portal can be found at \url{https://airquality.cpcb.gov.in/ccr/#/caaqm-dashboard-all/caaqm-landing}. This has stations all over India, and data can be extracted for the desired time period and study area.

\subsection{Meteorological Data}
Meteorological parameters, including rainfall, temperature, and wind speed, were incorporated into this study to enhance the accuracy of \noo~concentration predictions. These datasets were obtained from \gee.

\begin{figure}[tp]
    \centering
    \includegraphics[width=1.0\linewidth]{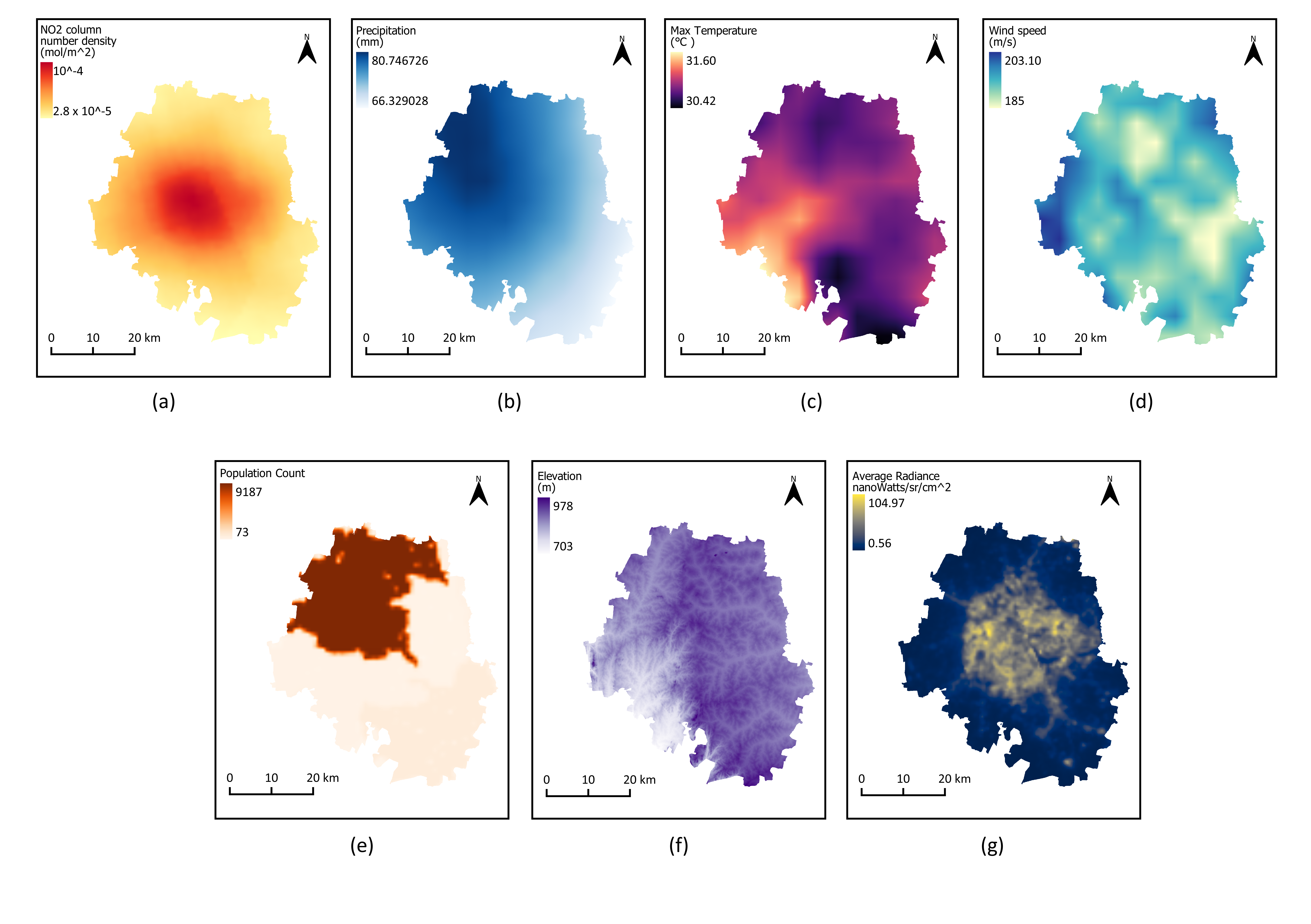}
    \caption{Yearly composite of driving factors over Bangalore for 2019. (a) TROPOMI Tropospheric \noo~Column Density, (b) Rainfall, (c) Temperature, (d) Wind Speed, (e) Population, (f) Elevation, (g) Nighttime Light Intensity}
    \label{fig:blr_drivingfactors}
\end{figure}

\begin{figure}[tp]
    \centering
    \includegraphics[width=1.0\linewidth]{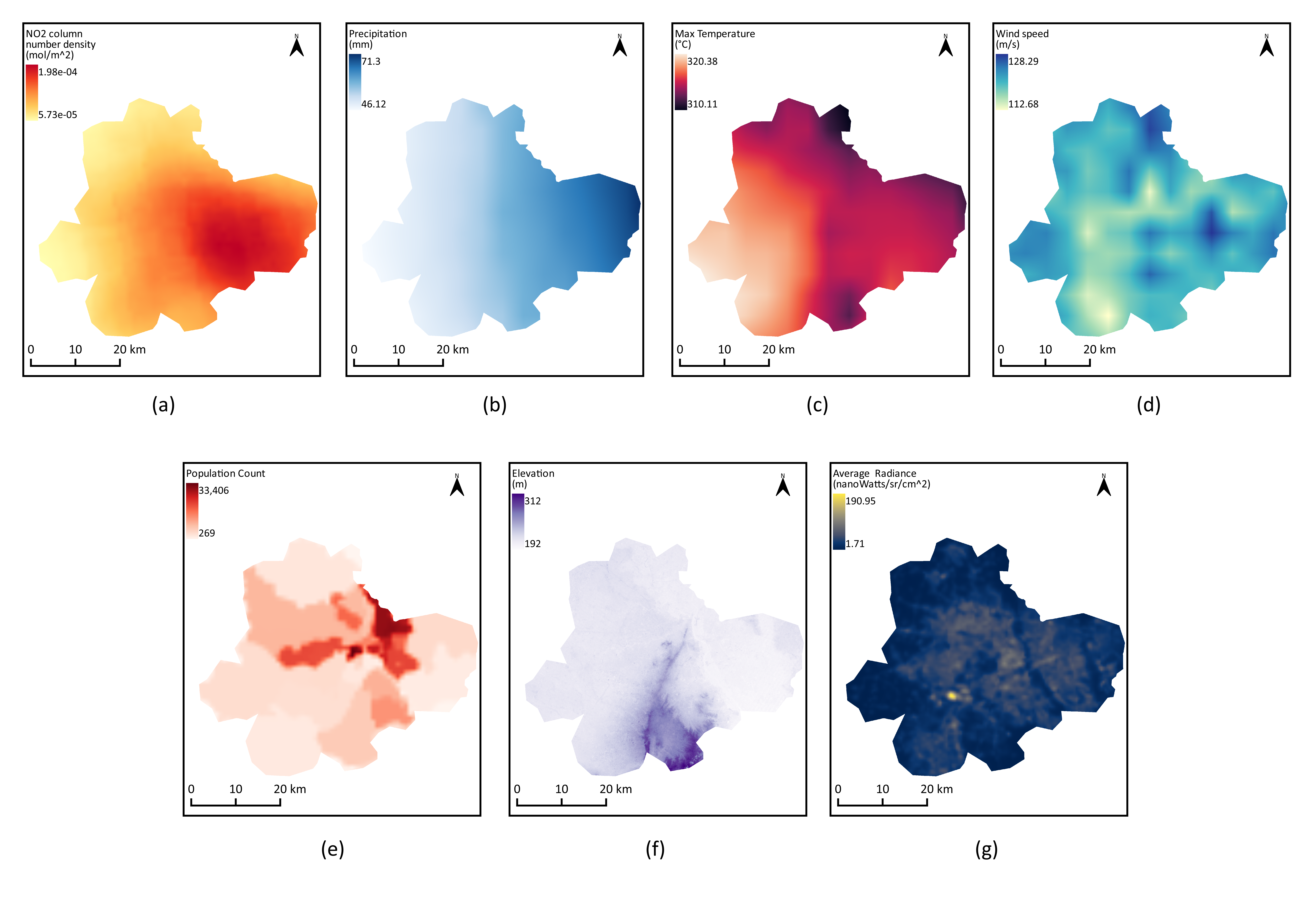}
    \caption{Yearly composite of driving factors over Delhi for 2019. (a) TROPOMI Tropospheric \noo~Column Density, (b) Rainfall, (c) Temperature, (d) Wind Speed, (e) Population, (f) Elevation, (g) Nighttime Light Intensity}
    \label{fig:del_drivingfactors}
\end{figure}

Rainfall data was sourced from the JAXA Global Precipitation Measurement Mission (GPM) GSMaP v6 operational dataset. This dataset provides near real-time precipitation estimates at a spatial resolution of $0.1^\circ$ and a temporal resolution of 1 hour. The yearly average of 2019 can be seen in Figures~\ref{fig:blr_drivingfactors}(b) and~\ref{fig:del_drivingfactors}(b). The GPM GSMaP product is particularly suitable for high-resolution rainfall analysis, offering global coverage and frequent updates, which are essential for capturing precipitation patterns relevant to atmospheric \noo~dynamics.

Temperature and wind speed data were derived from the TerraClimate dataset, hosted on \gee~as IDAHO EPSCoR TERRACLIMATE. This dataset provides monthly climate data, including temperature and wind speed, at a spatial resolution close to 4 km. The yearly composite of 2019 temperature and wind speed can be seen in Figures~\ref{fig:blr_drivingfactors}(c) and~\ref{fig:blr_drivingfactors}(d), respectively, for Bangalore, and similarly, Figures~\ref{fig:del_drivingfactors}(c) and~\ref{fig:del_drivingfactors}(d) for Delhi. TerraClimate incorporates high-quality ground observations and advanced climate reanalysis data, ensuring accuracy and consistency across global scales.

\subsection{Population Data}
Population data were utilized in this study to account for human activity and its influence on ambient \noo~concentrations. The population dataset was obtained from the Gridded Population of the World, Version 4 (GPWv4): Population Count, Revision~11, hosted on \gee.

This dataset provides estimates of population counts at a spatial resolution of approximately 1 km (30 arc-seconds), aggregated to represent populations for specific years. It is derived from national population censuses and incorporates administrative boundary data to ensure high spatial accuracy. The dataset used in this study represents population distribution at a global scale, enabling the analysis of human density and its correlation with NO\textsubscript{2} emissions. Since this dataset has yearly granularity, fixed values were used for a single year of study. The 2019 data is visualized in Figures~\ref{fig:blr_drivingfactors}(e) and~\ref{fig:del_drivingfactors}(e).

\subsection{Elevation Data}
Elevation data were included in the study to account for the impact of topography on \noo~dispersion and concentration. The elevation dataset was sourced from the CGIAR-CSI SRTM 90m Digital Elevation Database, Version 4, available on \gee.

This dataset provides global elevation data at a spatial resolution of 90 meters, derived from the Shuttle Radar Topography Mission (SRTM). It covers latitudes between $60^\circ$N and $56^\circ$S, offering a highly accurate representation of the Earth's surface topography. Since this data is also constant for the single-year study we are doing, constant values were used. These visualizations can be seen in Figures~\ref{fig:blr_drivingfactors}(f) and~\ref{fig:del_drivingfactors}(f).

\subsection{Nighttime Light Data}
Nighttime light intensity was used as a proxy for human activities and urbanization levels in this study, contributing to the analysis of their impact on NO2 concentrations. The data was obtained from the NOAA VIIRS DNB Monthly Composite Version 1 (VCMCFG) dataset, hosted on \gee.

This dataset provides monthly composites of nighttime lights captured by the Visible Infrared Imaging Radiometer Suite (VIIRS) Day/Night Band (DNB). The dataset is of global coverage with a spatial resolution of 15 arc-seconds (~500 meters), ensuring detailed insights into light emissions. It uses radiance-calibrated measurements, filtered to remove natural light sources like moonlight and auroras, and focuses solely on human-made light emissions whose visualizations can be seen in Figures~\ref{fig:blr_drivingfactors}(g) and~\ref{fig:del_drivingfactors}(g). 

\section{Method} \label{sec:method}
Our proposed tool, \caqv, is designed as an interactive \ml~sandbox that facilitates the development and evaluation of air quality prediction models. The tool implements a data processing workflow that enables users to interact with each stage of the workflow, allowing data selection, model customization, and real-time visualization of results, thereby promoting transparency, experimentation, and adaptability in urban air quality modeling.

\subsection{Workflow}

The end-to-end data processing workflow for \caqv~implementation involves three key stages (Figure~\ref{fig:workflow}): data curation, dataset preparation, and model training.

\subsubsection{Stages 1 and 2: Data Curation and Dataset Preparation}
\begin{figure}[t]
  \centering
    \includegraphics[width=0.9\textwidth]{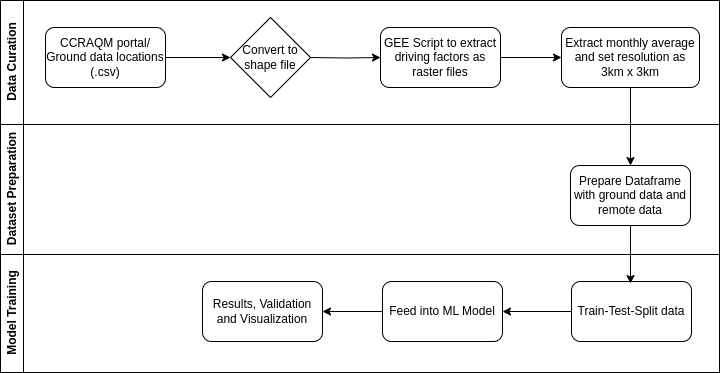}
    \caption{Data processing workflow of \caqv~using Machine Learning models}
    \label{fig:workflow}
\end{figure}

To predict the surface \noo, we curated the dataset with the driving factors mentioned in Section~\ref{sec:datasets}. We extracted driving factors as raster files for each month in each study area and then extracted the raster values to build a numerically valued dataset. For this, we used the Python library rasterio. All the datasets described above have different spatio-temporal resolutions; therefore, it is essential to preprocess them to a uniform resolution prior to model training. We decided to go ahead with a 3km$\times$3km grid for both study areas as this was the most reasonable given the resolutions of all datasets and the size of cities (Bangalore 741 km\textsuperscript{2} and Delhi 247 km\textsuperscript{2}). We extracted all raster files from \gee~scripts, setting the export resolution to 3000m$\times$3000m, thus ensuring uniform spatial resolution data. For the ground data, we took the monthly average for each station. Figure~\ref{fig:workflow} shows the overall workflow of data processing.

\subsubsection{Stage 3: Machine Learning Model Training}\label{sec:method-ml}
Our independent variables consist of TROPOMI NO\textsubscript{2}, temperature, population, windspeed, rainfall, elevation, and VIRS. Using these, we trained various \ml~models to predict the ground-level NO\textsubscript{2}. We used four models for both the Bangalore and Delhi study areas to demonstrate model comparison:
\begin{enumerate}
    \item Linear Regression
    \item Random Forest along with GridSearch for optimization
    \item Support Vector Machine
    \item Gradient Boosting Regressor along with GridSearch for optimization
\end{enumerate}
For each model, we performed basic preprocessing by removing null values and applying a 70:30 train-test split for model training. To optimize performance, we utilized GridSearchCV from scikit-learn to tune hyperparameters for Random Forest and Gradient Boosting models, ensuring better predictive accuracy. These models have been integrated into our sandbox environment, allowing users to experiment with and compare different \ml~approaches for air quality prediction.

\subsection{Machine Learning Sandbox and Visualization Tool}
\begin{figure}[t]
  \centering
  \includegraphics[width=1.0\textwidth]{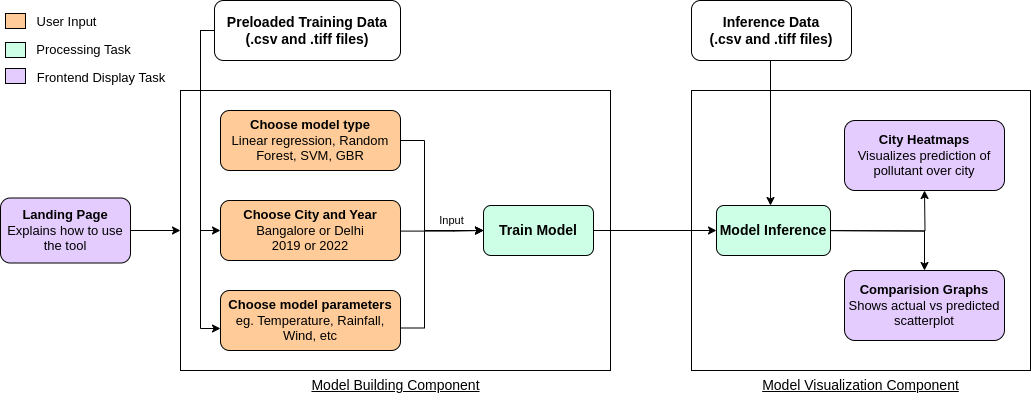}
  \caption{Workflow of the Visualization Tool}
  \label{fig:gui}
\end{figure}

To facilitate the exploration and application of \ml~models for air quality prediction, we developed an interactive sandbox environment that enables users to train, compare, and visualize NO$_2$ predictions across urban areas. This tool allows for flexible experimentation with different \ml~models and datasets, making it adaptable to various cities and pollutants. The interface provides an intuitive visualization of model outputs, aiding in the interpretation of spatial pollution patterns. Figure~\ref{fig:gui} illustrates the workflow of user interactions implemented by the tool.

\subsubsection{User Inferface Design}
The Graphical User Interface (GUI) is designed as a clean two-panel view, as shown in Figure~\ref{fig:model-building}. The two panels are then used for comparative analysis. Such an analysis involves adaptively setting up model parameters and inputs for training. The visualizations are placed appropriately below the parameter setting so that vertical scrolling of the page will provide the context and information within the self-contained panel.

\subsubsection{User Interactions for Model Building}
The tool has been designed with a simple and intuitive user interface. On the landing page, we have displayed basic guidelines and information on how to use and extend it for a custom use case, like different cities and different pollutants.

\begin{figure}
    \centering
    \includegraphics[width=1.0\linewidth]{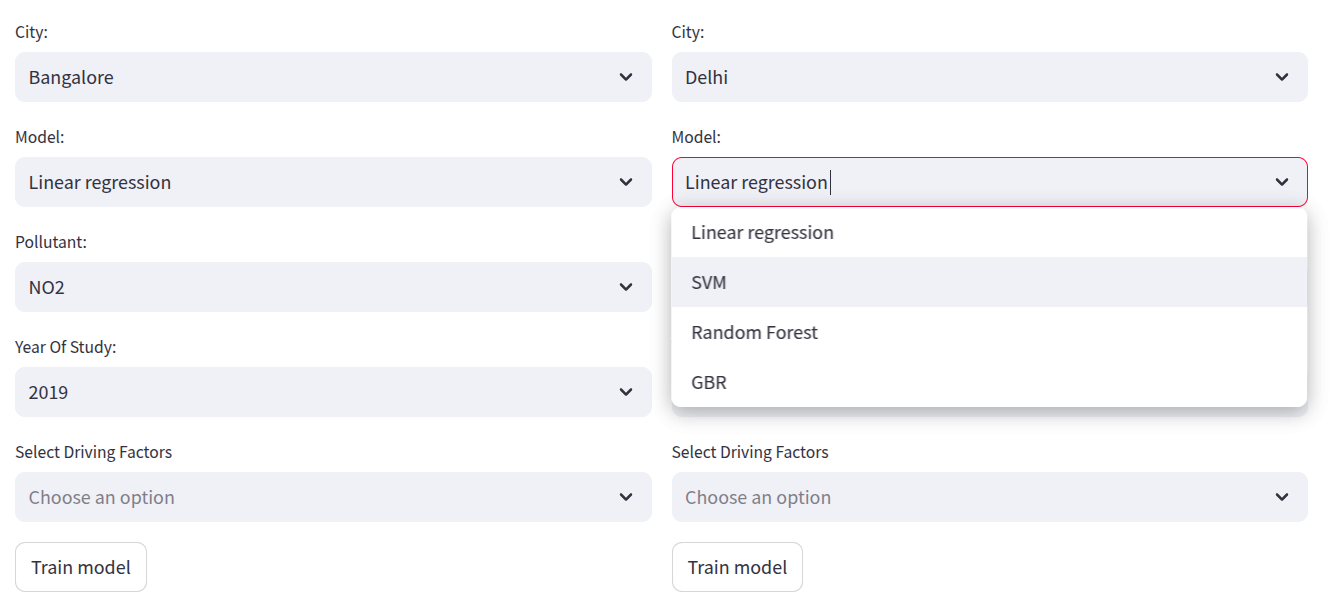}
    \caption{User can select parameters to build a custom model on \caqv}
    \label{fig:model-building}
\end{figure}

The tool allows users to customize the analysis by selecting the city, year, pollutant, model type, and a subset of driving factors to be used, as shown in Figure~\ref{fig:model-building}.

These user-defined parameters serve as inputs, guiding the model training process.  The training dataset, preloaded for the specified city, year, and pollutant, is filtered to include only the selected driving factors as independent variables, and the model is trained only on these. Once the model is trained, the next step is to visualize the output over the area of interest.

\subsubsection{Methods and User Interactions for Visualizations}

To use the trained model for inference, a set of input points is required. These inputs consist of the same user-selected driving factors used during training. The city is then divided into a uniform grid of points, defined by their latitude and longitude, generated through a grid-like mapping with horizontal and vertical intervals determined by the city's perimeter. For each point on the grid, the yearly composite of the selected driving factors is extracted from the raster files and fed into the trained model. This process predicts ground-level \noo~values at each grid point, which is then visualized across the entire study area.

Since the input is a collection of discrete points, the output will not be a continuous series of values. Also, owing to the limited number of ground monitoring stations, the input points tend to be sparse. Hence, smoothly visualizing the prediction in an area based on sparse discrete input points requires modifications by design. 

Initially, we explored using Kernel Density Estimation (KDE) to visualize the sparse points as a continuous surface, aiming to create a smooth representation of the spatial distribution of \noo~concentrations. KDE is well-suited for interpolating values across a study area, making it possible to depict patterns more effectively in regions with fewer data points.

However, the computational intensity of KDE  significantly slowed down the responsiveness of the web application tool, especially when applied to large datasets or high-resolution grids. Given the importance of real-time interactivity for a responsive web application, this approach was deemed impractical, and we decided to focus on grid-based predictions and visualization instead.

So to improve the user experience and maintain a seamless feel when visualizing \noo~predictions, we implemented a zoom factor limitation on the map. This approach ensures that the output is displayed as a continuous block, effectively masking the sparsity of the input data points derived from the grid mapping described earlier. By restricting excessive zooming, the visualizations remain intuitive and aesthetically cohesive, preventing the granularity of the grid points from disrupting the overall perception of the predicted spatial patterns.

After the model is trained, the yearly composite of other selected driving factors are used as inference, and the predicted output, \ie~ground-level  values of \noo, is visualized on the map. These yearly values are extracted in the same manner as described in Section~\ref{sec:datasets}, but instead of a monthly average, a yearly average is computed for interactive visualization, with more focus on the tool as a proof of concept. Further computational optimizations in the future work are planned to increase the temporal granularity in our analysis.

Users can see how the model performed in terms of validation metrics, which include R\textsuperscript{2} Score, Mean Absolute Error (MAE), Mean Squared Error (MSE), Mean Absolute Percentage Error (MAPE), and Root Mean Square Error (RMSE).

\subsubsection{Implementation}
The user interface of \caqv~was developed using Python’s \texttt{Streamlit} library, which enables rapid prototyping and deployment of interactive web applications with minimal overhead. \texttt{Streamlit} was chosen for its ease of integration with data science workflows and native support for interactive visualizations.

\begin{figure}[tp]
    \centering
    \includegraphics[width=0.9\textwidth]{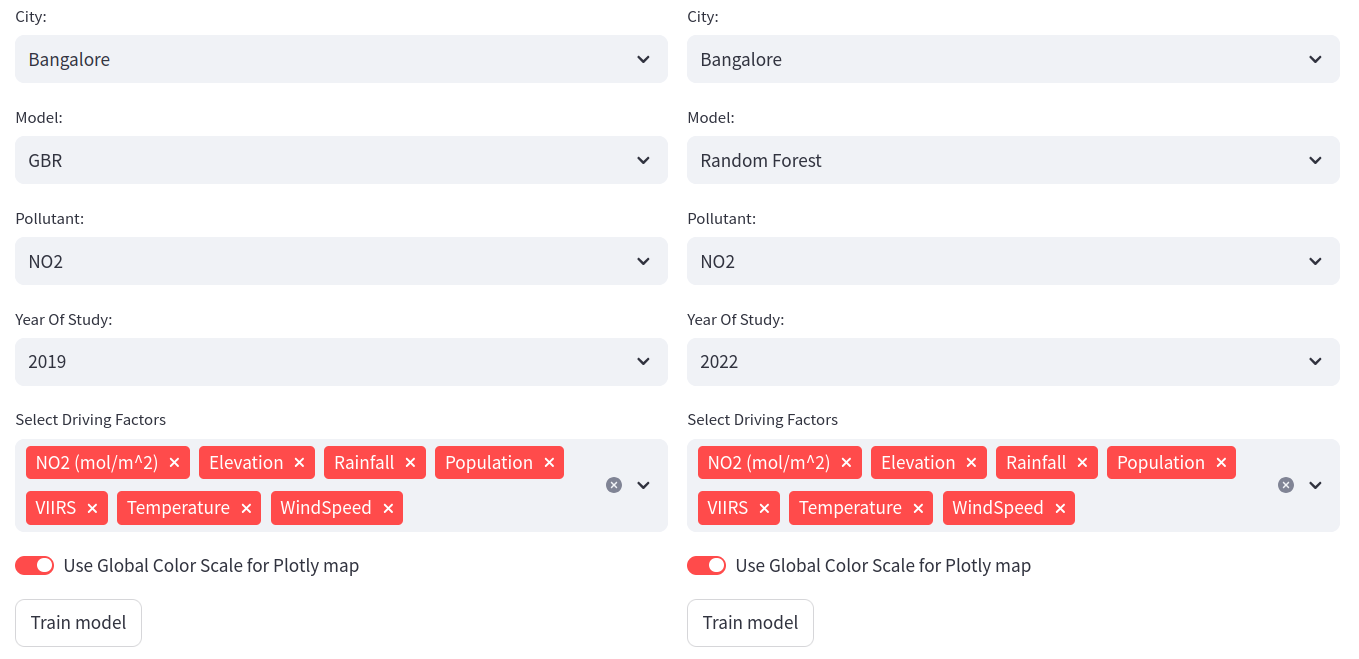}
    \par\vspace{1mm}
    \caption*{(a) User input for the case study}

    \vspace{2mm}

    \includegraphics[width=0.9\textwidth]{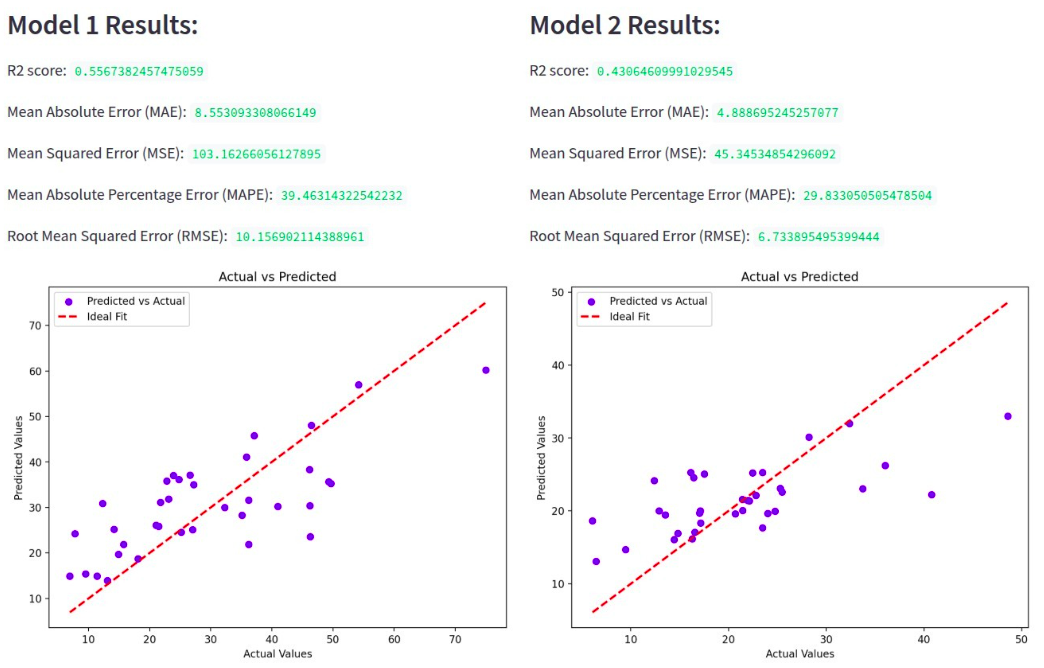}
    \par\vspace{1mm}
    \caption*{(b) Performance metrics and prediction results generated by \caqv}

    \caption{GUI setup and results of intra-city case study for Bangalore in 2019 and 2022}
    \label{fig:blr-study-1}
\end{figure}

For plotting geospatial data and prediction outputs, we used Python-based libraries \texttt{Folium} \citep{folium} and \texttt{Plotly} \citep{plotly}. \texttt{Folium} produces tile-based maps that facilitate detailed spatial overlays and contextual analysis, thereby providing users with multiple visualization perspectives. \texttt{Plotly} generates interactive, data-driven maps enabling dynamic exploration of prediction results. Examples of \texttt{Folium} and \texttt{Plotly} can be seen in Figure~\ref{fig:blr-study-1}(c) and Figure~\ref{fig:blr-study-1}(d) respectively. There are slight differences in both implementations. Since \texttt{Plotly} uses a dynamically changing legend as per the data, we included a toggle widget (Figure~\ref{fig:blr-study-1}(a)) that allows the user to switch the dynamic scale off and use a fixed global scale like the \texttt{Folium} map is using.

The backend processing, including model training and evaluation, leverages libraries such as \texttt{scikit-learn}~\citep{scikit-learn} for \ml, \texttt{Pandas}~\citep{reback2020pandas} for data handling, and \texttt{NumPy}~\citep{harris2020array} for numerical operations. This technology stack ensures that \caqv~remains lightweight, extensible, and accessible to both technical and non-technical users.

The pilot version of the tool was developed and tested on a local workstation running Ubuntu 22.04, equipped with an Intel Core i5 processor and 16 GB of RAM. While a GPU was not necessary for training the traditional \ml~models used in this version, the tool is designed to be compatible with more computationally intensive environments if extended to deep learning or larger-scale datasets. This setup was sufficient for training models on multi-year air quality data and delivering responsive, real-time visualizations through the interface.

\subsubsection{Processing Steps for Configurability of \caqv}
\caqv~can be configured, by design, for any urban area of interest and pollutant by following this sequence of data processing steps. The sample notebook mentioned below will help in understanding the expected intermediate outcomes. However, configuring the tool to change the preloaded data will require the user to have a basic level of Python coding skills.

\begin{enumerate}  
    \item \textbf{Download Ground Truth Station Data}:  
      Retrieve station data from the CCRAQM portal (for regions in India) or any similar source of field measurements of pollutants, that will be used as ground truth for the \ml~models.
      
    \item \textbf{Create CSV with Station Details}:  
    Prepare a CSV file containing station names along with their corresponding UTM (Universal Transverse Mercator) coordinates.  

    \item \textbf{Convert CSV to Shapefile}:  
    Convert the CSV file into a shapefile using Geographic Information System (GIS) software such as QGIS. This shapefile is necessary for spatial data processing.  

    \item \textbf{Extract Driving Factor Data}:  
    Utilize Google Earth Engine or any other relevant source to extract driving factor data (e.g., meteorological and land-use variables) in Tagged Image File Format (TIFF) format for the monitoring stations based on the shapefile.  

    \item \textbf{Dataset Preparation in Notebook}:  
    Load the collected data into a notebook or code file and execute the dataset preparation steps. The notebook used for the pilot implementation is available in the GitHub repository of this work, \url{https://github.com/GVCL/CityAQVis}.

    \item \textbf{Download Processed Data}:  
    Export and download the processed dataset, including the extracted driving factors, as a CSV file.  

    \item \textbf{Manually Add Ground Data}:  
    Append monthly averages of \noo~or other pollutant concentration values from the monitoring stations to the corresponding rows in the CSV file.    

    \item \textbf{Integrate Data with Web Application}:  
    To integrate the model with a web-based visualization tool, place the processed data in the designated application directory and update the corresponding file paths in the application’s codebase.  

    \item \textbf{Visualize Predictions}:  
    \begin{itemize}  
        \item Obtain a shapefile for the target area.  
        \item Extract driving factor data for inference from TIFF files. For simplicity, a yearly composite was used for this paper; any data range can be used for a custom purpose. 
        \item Generate a uniform grid of latitude-longitude points over the study region.  
        \item Predict pollutant levels at each grid point using the trained model.    
    \end{itemize}  

\end{enumerate} 

\begin{figure}[tp]
    \centering
    \includegraphics[width=0.85\textwidth]{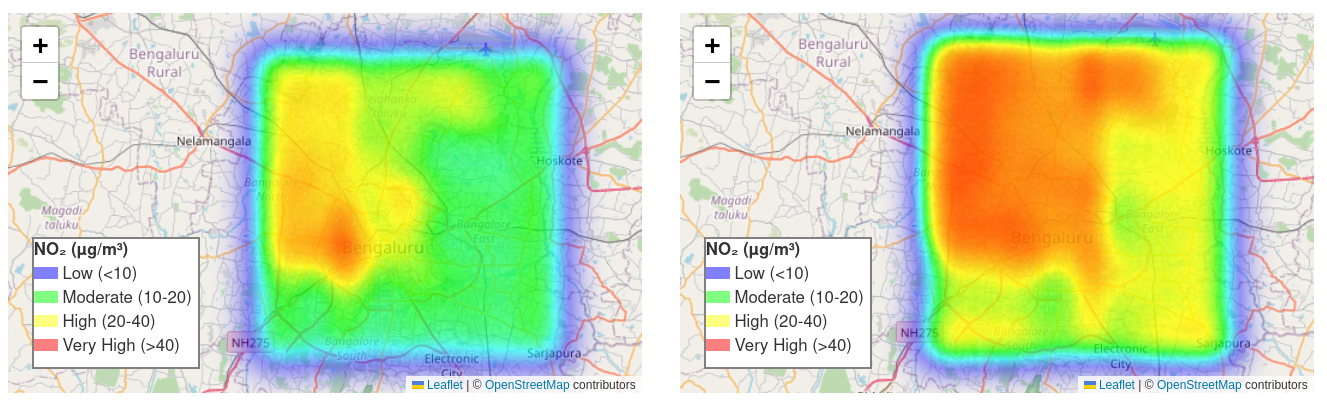}
    \par\vspace{1mm}
    \caption*{(a) \texttt{Folium} visualization of predicted \noo~values in 2019(L) and 2022(R)}

    \vspace{2mm}

    \includegraphics[width=0.85\textwidth]{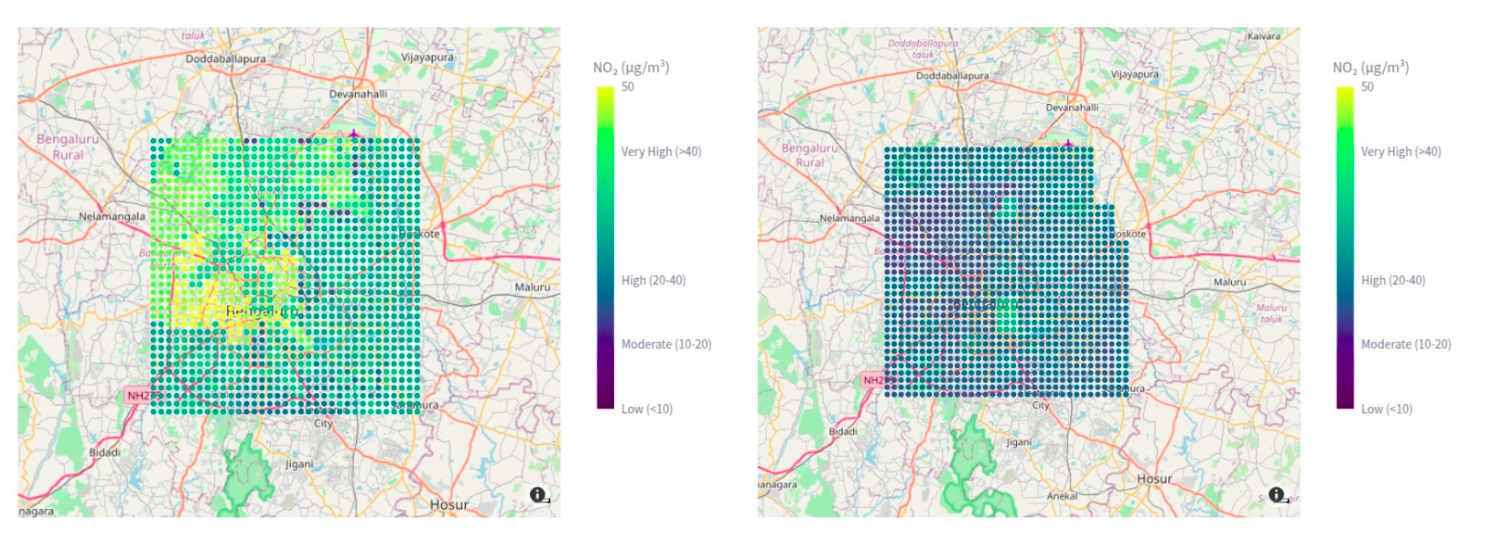}
    \par\vspace{1mm}
    \caption*{(b) \texttt{Plotly} visualization of predicted \noo~values in 2019(L) and 2022(R)}

    \caption{Intra-city case study visualizations for Bangalore in 2019 and 2022}
    \label{fig:blr-study-2}
\end{figure}

\subsection{Case Study to Demonstrate Tool Usage}
We have mainly performed two case studies to demonstrate the overall scope and capabilities of the tool. The first study is conducted over a single city but across different time periods, and the second one compares two cities over the same time period. In both studies, we compare the results of the best \ml~model that worked for the input dataset of the study area.

\subsubsection{Intra-City Comparative Analysis}
To evaluate the capabilities of \caqv, we conducted a comparative analysis of \noo~levels in Bangalore for the years 2019 and 2022. This is a comparison of the air quality before and after COVID-19, which is studied widely in literature (Section~\ref{sec:related_work}). For the 2019 analysis, we employed the Gradient Boosting Regressor model with all available driving factors mentioned in Section~\ref{sec:datasets}, while for 2022, we utilized the Random Forest model with the same set of factors, as illustrated in Figure~\ref{fig:blr-study-1}(a).

Subsequent to model training, we generated the performance metrics and visualizations for both years to facilitate direct comparisons of the model's performance. Figure~\ref{fig:blr-study-1}(b) presents key validation metrics, including R$^2$ scores, Mean Absolute Error (MAE), Mean Squared Error (MSE), Mean Absolute Percentage Error (MAPE), and Root Mean Squared Error (RMSE), alongside actual versus predicted graphs. The results indicate that while the 2019 model achieved a higher R$^2$ score, suggesting better overall explanatory power, the 2022 model exhibited lower error values (MAE, MSE, RMSE), indicating improved prediction accuracy.

Additionally, spatial visualizations of the predicted \noo~concentrations for both years, shown in Figure~\ref{fig:blr-study-2}(a) and Figure~\ref{fig:blr-study-2}(b), reveal a significant shift in pollution distribution. In 2019, high \noo~concentrations were more localized, particularly in the upper-left region of the city, indicating concentrated emission sources. By 2022, the intensity appears more diffused across the city, with a reduction in peak \noo~values. The predicted decrease in \noo~concentrations over Bangalore between 2019 and 2022 is in agreement with previous studies that have documented significant reductions during COVID-19 lockdown periods due to decreased vehicular emissions and industrial activities~\citep{suthar2022four}.

\begin{figure}[tp]
    \centering

    \includegraphics[width=0.9\textwidth]{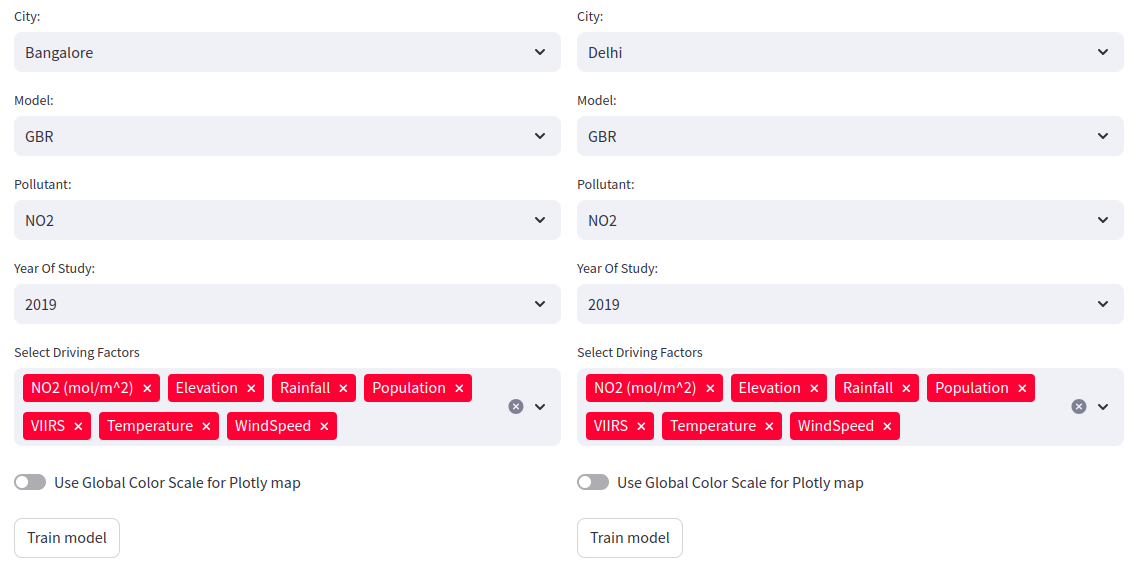}
    \par\vspace{1mm}
    \caption*{(a) User input for the case study}

    \vspace{2mm}

    \includegraphics[width=0.9\textwidth]{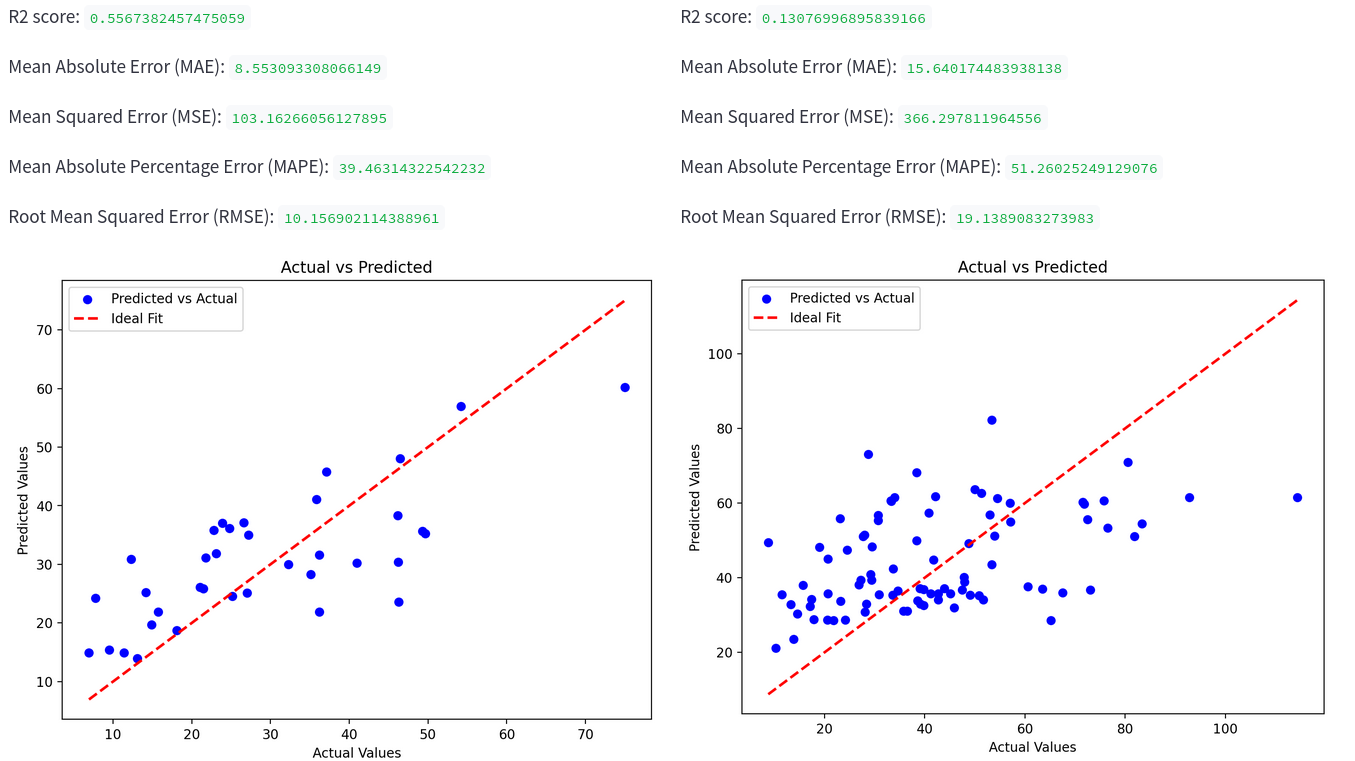}
    \par\vspace{1mm}
    \caption*{(b) Performance metrics and prediction results generated by \caqv}

    \caption{GUI setup and results of inter-city case study for Bangalore and Delhi in 2019.}
    \label{fig:blrDel-study-1}
\end{figure}

\begin{figure}[tp]
    \centering

    \includegraphics[width=0.85\textwidth]{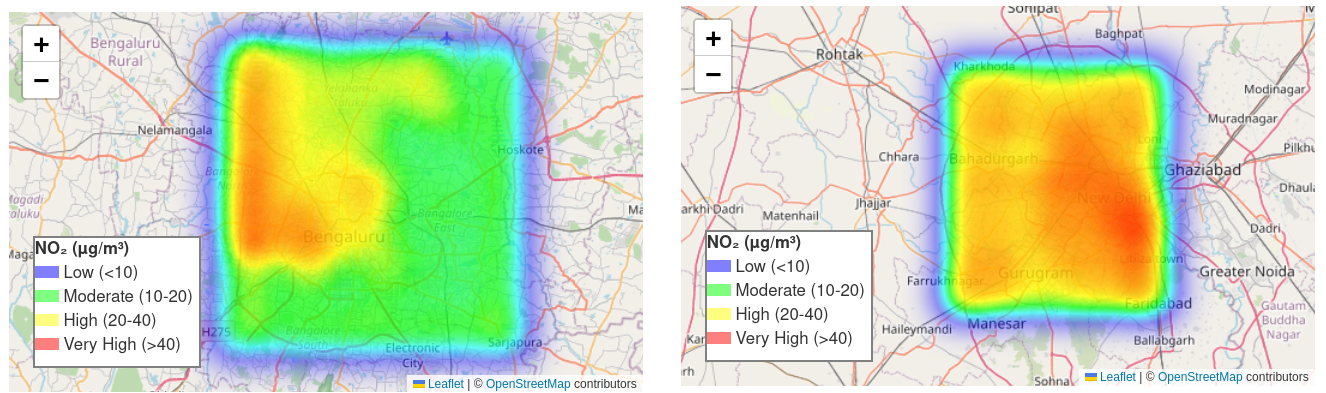}
    \par\vspace{1mm}
    \caption*{(a) \texttt{Folium} visualization of predicted \noo~values in Bangalore(L) and Delhi(R)}

    \vspace{2mm}

    \includegraphics[width=0.85\textwidth]{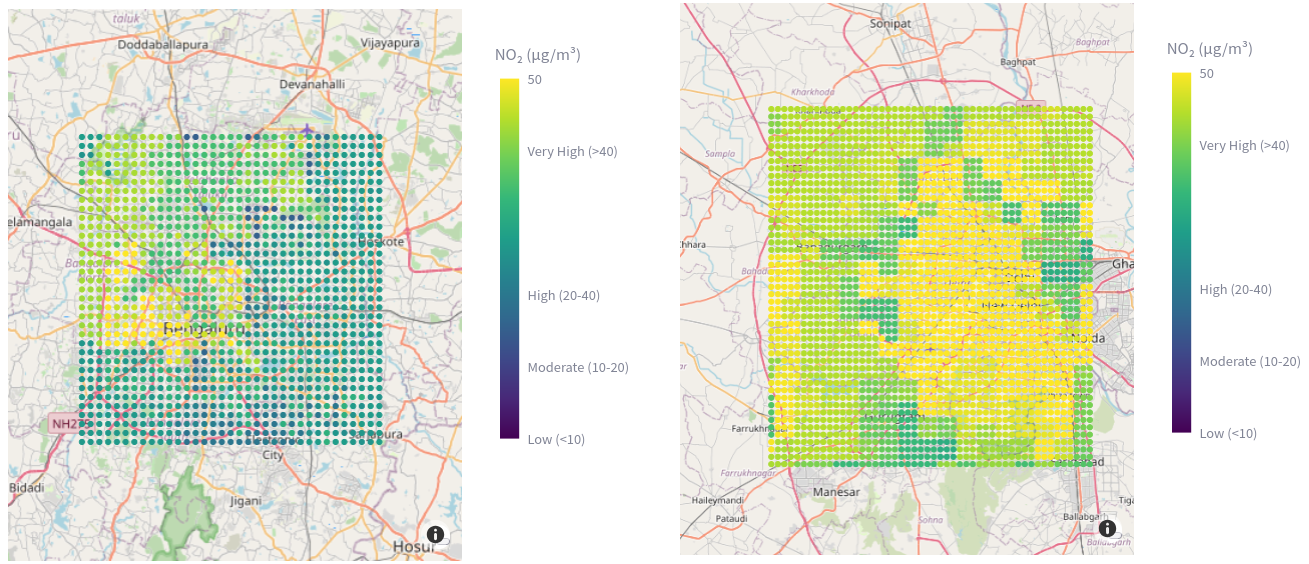}
    \par\vspace{1mm}
    \caption*{(b) \texttt{Plotly} visualization of predicted \noo~values in Bangalore(L) and Delhi(R)}

    \caption{Inter-city case study visualizations for Bangalore and Delhi in 2019.}
    \label{fig:blrDel-study-2}
\end{figure}

\subsubsection{Inter-City Comparative Analysis}

\caqv~also facilitates comparative analyses between different cities. To demonstrate this capability, we conducted a study comparing \noo~levels in Bangalore and Delhi for the year 2019. The same set of driving factors used in the previous experiment was employed for both cities, with the Gradient Boosting Regressor selected as the predictive model, as illustrated in Figure~\ref{fig:blrDel-study-1}(a).

Following model training, performance metrics and visualizations were generated to assess model effectiveness in both urban settings. Figure~\ref{fig:blrDel-study-1}(b) presents key validation metrics, including R², MAE, MSE, MAPE, and RMSE, alongside actual versus predicted graphs. The model yielded better performance for Bangalore, achieving higher R$^2$ and lower error values, suggesting more stable prediction behavior compared to that of Delhi.

Additionally, Figure~\ref{fig:blrDel-study-2}(a) and Figure~\ref{fig:blrDel-study-2}(b) present spatial visualizations of the predicted \noo concentrations for each city. These maps highlight stark contrasts. While Bangalore exhibited more localized and moderate concentration zones, Delhi displayed significantly higher \noo~levels distributed more widely across the urban area. This observation aligns with existing literature on Delhi’s chronic air pollution challenges, further validating the tool's visual comparative air quality analysis across different urban regions.

\subsubsection{Results and Validation} \label{sec:results}
\begin{table}[tp]
\centering
\caption{Performance Metrics for Different \ml~Models - Bangalore and Delhi in 2019}
\begin{tabular}{lccccc}
\hline
\textbf{Model}
& \textbf{R\textsuperscript{2} Score} & \textbf{MAE}
& \textbf{MSE} & \textbf{MAPE (\%)} & \textbf{RMSE}
\\ \hline

\multicolumn{6}{c}{\bf Bangalore, India}
\\\hline

Random Forest
& 0.483  & 9.11
& 120.26 & 43.75 & 10.97
\\ 

Linear Regression
& 0.371 & 9.78
& 146.29 & 48.77 & 12.10
\\ 

\textbf{Support Vector Machine}
& \textbf{0.672} & \textbf{7.19}
& \textbf{76.38} & \textbf{30.90} & \textbf{8.74}
\\

Gradient Boosting Regression
& 0.573 & 8.34
& 99.49 & 39.13 & 9.97
\\ \hline

\multicolumn{6}{c}{\bf Delhi, India}
\\ \hline

\textbf{Random Forest}
& \textbf{0.410} & \textbf{12.71}
& \textbf{289.35} & \textbf{38.44} & \textbf{17.01}
\\

Linear Regression
& 0.190 & 15.17
& 397.20 & 47.96 & 19.93
\\

Support Vector Machine
& 0.308 & 13.65
& 339.05 & 39.44 & 18.41
\\

Gradient Boosting Regression
& 0.362 & 13.31
& 312.74 & 39.95 & 17.68
\\ \hline
\end{tabular}
\label{tab:model_performance}
\end{table} 

In this section, we present the performance metrics of the ML models mentioned in Section \ref{sec:method-ml} for predicting ground-level/surface \noo~concentrations in Bangalore and Delhi for the year 2019. The results showcase the variability in model effectiveness across different urban scenarios, highlighting the capabilities and limitations of each method. The performance metrics for Bangalore and Delhi are presented in Table \ref{tab:model_performance}. 
\\
\\Our key findings are:
\begin{itemize}
\item \textbf{Bangalore}: The \textbf{Support Vector Machine} model outperformed the others with an R\textsuperscript{2} score of 0.672 and the lowest MAE of 7.19, indicating superior predictive capability for this dataset.
\item \textbf{Delhi}: The metrics include R\textsuperscript{2} Score, MAE, MSE, MAPE, and RMSE. The \textbf{Random Forest Model} performed the best in terms of all the metrics.
\end{itemize}

These results highlight the varying effectiveness of models in predicting ground-level \noo~concentrations across different urban environments. In Delhi, the Random Forest model demonstrated superior performance, achieving the highest R$^2$ score and the lowest error metrics among the models tested. This can be attributed to its ability to handle complex interactions and non-linear relationships between variables, which are prevalent in Delhi due to its dense urban structure and diverse pollution sources. In contrast, the Support Vector Machine (SVM) model performed best in Bangalore, achieving the highest R$^2$ score and lowest MAE. The city's moderate pollution levels and distinct meteorological and topographical conditions may favor SVM's strengths in finding optimal boundaries in the feature space.

\section{Software Files} \label{sec:software}
The source code for \caqv~is publicly available in a GitHub repository, which can be accessed at \url{https://github.com/GVCL/CityAQVis}. This repository contains all the necessary scripts for data preprocessing, model inference, and visualization. Furthermore, the repository includes the datasets utilized in our case study, along with detailed instructions to reproduce the results presented in this work, ensuring the transparency and reproducibility of our research. A live demonstration of the tool is accessible at \url{https://cityaqvis-gqj9xf7aqvnj3wkiljvcyf.streamlit.app/}. It is important to note that the application may experience a slight delay in loading as it is deployed on a free community cloud service.

The \caqv~tool was developed and tested on a system equipped with an Intel Core i5 (11th Gen) processor, 16 GB of RAM, and running the Ubuntu 22.04 LTS operating system. All computations were performed on the central processing unit (CPU), without the use of a dedicated graphics processing unit (GPU). The software environment was constructed using Python 3.10, with key libraries including \texttt{NumPy}, \texttt{Pandas}, \texttt{Scikit-learn}, \texttt{Plotly}, \texttt{Folium}, and \texttt{PyTorch}. The lightweight nature of \caqv~allows for its efficient operation on standard laptop computers, making it accessible for practical research and policy-making workflows.

\section{Discussion and Conclusion} \label{sec:conclusion}
Our findings from the case study suggest that no single model universally outperforms others and that the effectiveness of a model is strongly dependent on the specific characteristics of the study area. Thus, \caqv~is designed to provide users with the flexibility to experiment with different models and datasets. This adaptability is particularly valuable for urban planners and environmental researchers who require tailored solutions for different cities and pollutants. By allowing interactive visualization of both model predictions and input variables, the tool fosters informed decision-making and a deeper understanding of urban air quality dynamics.

It is important to distinguish our approach from the hotspot analysis tools available in commercial GIS software such as ArcGIS Pro \citep{esri2023arcgis}. ArcGIS Pro supports both kernel density estimation (KDE) and hotspot analysis based on the Getis-Ord Gi* statistic. KDE generates continuous density surfaces that highlight regions of high event intensity, whereas hotspot analysis statistically identifies clusters of high or low values unlikely to occur by chance. In contrast, \caqv~focuses on prediction-driven visualizations using \ml models, enabling users to generate spatial distributions of pollutant concentrations rather than density surfaces alone. Our case study also highlights the challenges of working with sparse point data from ground air monitoring stations, a common scenario in developing regions. While ArcGIS Pro provides sophisticated handling of dense spatial datasets, its performance with sparse monitoring networks often requires interpolation or resampling. By integrating sparse station data with multi-source satellite and meteorological datasets, \caqv~is specifically tailored for such contexts.

Another key distinction lies in usability and flexibility. Traditional platforms such as ArcGIS Pro are highly feature-rich but generally optimized for the analysis of a single dataset at a time, where comparative workflows often require opening multiple instances or manually synchronizing views. \caqv, by contrast, embeds machine learning and visualization into an analytic workflow explicitly designed for comparison tasks, allowing users to evaluate multiple models, time periods, or cities within a single interactive interface. Furthermore, as a Python-based tool, \caqv~is readily accessible to developers and researchers to customize or extend its functionality, offering a level of adaptability that proprietary GIS software typically does not provide.

In conclusion, \caqv~provides a user-friendly GUI to explore the interaction between data sources, ML models, and urban air quality. By allowing users to extend the tool to their cities, it encourages widespread adoption and customization for local contexts. Our future work will focus on integrating additional data sources, such as vehicular traffic patterns and industrial emission inventories, to further improve the model accuracy. We also plan to enhance the interaction design by synchronizing panels to support frequently used comparative analysis tasks, further streamlining workflows for analysts working on geospatial air quality data.  

\bibliographystyle{alpha}
\bibliography{papers_pollution}
\end{document}